\documentclass[12pt]{article}

\usepackage[sectionbib]{natbib}
\usepackage{caption}
\usepackage{float}
\usepackage{graphicx,psfrag,epsf}
\usepackage{times}
\usepackage{amsmath,amsfonts}
\usepackage{natbib}
\usepackage{enumitem}
\usepackage{color}

\usepackage{amsmath}
\usepackage{graphicx,psfrag,epsf}
\usepackage{enumerate}
\usepackage{subfig}
\usepackage{epstopdf}
\usepackage{verbatim}
\usepackage{url} 
\usepackage{epstopdf}
\usepackage{multirow}
\usepackage{color}
\usepackage{mathrsfs}
\usepackage{indentfirst}
\usepackage{algorithm}
\usepackage{algpseudocode}
\usepackage{subfig}
\usepackage{amssymb,amsfonts,amsthm,amsmath}
\graphicspath{{"./images/"}}

\DeclareMathOperator*{\argmin}{arg\,min}
\DeclareMathOperator*{\argmax}{arg\,max}

\newcommand{\bm}[1]{\mbox{\boldmath$ #1 $\unboldmath}}
\theoremstyle{definition}

\begin{document}

\def\spacingset#1{\renewcommand{\baselinestretch}%
{#1}\small\normalsize} \spacingset{1}


\begin{center}
    {\Large\bf Robust Experimental Designs for Model Calibration}\\
\vspace{.5in}
Arvind Krishna \\
\vspace{.1in}
V. Roshan Joseph \\
\vspace{.1in}
{\small H. Milton Stewart School of Industrial and Systems
Engineering,} \\
{\small Georgia Institute of Technology, Atlanta, GA 30332} \\
\vspace{.5in}
Shan Ba \\
\vspace{.1in}
{\small LinkedIn Corporation, Sunnyvale, CA 94085} \\
\vspace{.5in}
William A. Brenneman\\
\vspace{.1in}
William R. Myers\\
\vspace{.1in}
{\small Procter \& Gamble Company, Cincinnati, OH 45040} \\
\end{center}
\vspace{.4in}
\begin{abstract}
A computer model can be used for predicting an output only after specifying the values of some unknown physical constants known as calibration parameters. The unknown calibration parameters can be estimated from real data by conducting physical experiments. This paper presents an approach to optimally design such a physical experiment. The problem of optimally designing physical experiment, using a computer model, is similar to the problem of finding optimal design for fitting nonlinear models. However, the problem is more challenging than the existing work on nonlinear optimal design because of the possibility of model discrepancy, that is, the computer model may not be an accurate representation of the true underlying model. Therefore, we propose an optimal design approach that is robust to potential model discrepancies. We show that our designs are better than the commonly used physical experimental designs that do not make use of the information contained in the computer model and other nonlinear optimal designs that ignore potential model discrepancies. We illustrate our approach using a toy example and a real example from industry.
\end{abstract}

\noindent%

\newpage
\spacingset{1.45} 
\section{Introduction}
\label{2sec:intro}

A physical system can be explored or optimized by conducting experiments, but they can be expensive and time consuming. The whole field of experimental design in statistics focus on how to perform these experiments in an efficient way so that maximum information about the system can be obtained with minimum cost \citep{wu2011experiments}. Another way to reduce the experimental cost is to develop mathematical models that can mimic the physical system and explore the system through simulations \citep{santner2018space}.  These mathematical models can be very complex, such as a system of partial differential equations, which needs to be solved numerically. A computer implementation of the mathematical models is sometimes called computer models. Exploration using computer models is useful and can provide conclusive results only if they are good in representing the physical system. However, mathematical models and therefore, the computer models are only an approximation to the complex phenomenon that we are trying to explore in the physical system. Therefore, the computer simulations should only be used to assist the physical experimentation. In other words, physical experiments should always be performed to validate the computer models. This article examines how to perform a physical experiment when a computer model is available to the investigator.

A computer model can contain unknown parameters. Choosing them based on the physical experimental data can make the computer models closer to the reality. This approach is known as model calibration and the unknown parameters are often referred to as calibration parameters \citep{box1962}. Therefore, one possible approach to physical experiments is to design them in such a way that the calibration parameters can be estimated efficiently from data. Since the computer models are often nonlinear in the calibration parameters, one can use results from the nonlinear optimal design theory to design such experiments \citep{silvey1980optimal}. One of the major pitfalls of this approach is that the optimal design is overly dependent on the computer model and may not be capable of detecting possible violations. However, detecting such possible violations is at the core of the model calibration problem and is our main aim. Thus we need to design experiments based on the computer model, but it should not rely too much on the computer model and should be robust to the misspecifications of the model.

The importance of designing experiments robust to the model assumptions has long been recognized in the literature starting with \cite{box1959basis}. A good account of the follow-up works on their seminal paper and other related developments can be found in the review by \cite{chang199628}. Unfortunately, most of these approaches rely on an alternative class of possible models, the specification of which is non-trivial. Therefore, the research in this area has not led to practically implementable solutions other than in very simple settings such as adding a center point when fitting a plane, etc. Moreover, the literature seems to be scarce on model robust designs for nonlinear models. Nonlinear models add complexity to the problem because the optimal designs become functions of the unknown parameters. Robust designs can be developed by expressing the uncertainties in the unknown parameters through a prior distribution and using Bayesian optimal designs \citep{chaloner1995bayesian} or psuedo-Bayesian optimal designs \citep{dror2006robust}. Introducing model uncertainty also into this framework makes the problem much harder to solve. Furthermore, most of the existing works focus on simple nonlinear models such as a logistic regression model and not on the complex computer models that we are interested in. Computationally expensive computer models add another layer of complexity because they are known only in the places where the computer simulations have been performed. Therefore, we also need to entertain uncertainties arising due to the incomplete knowledge about the true computer model output.

We are not the first to look into the problem of designing physical experiments when computer models are available. Based on the Bayesian model calibration framework of \cite{kennedy2001bayesian}, optimal designs for both computer and physical experiments using integrated mean squared prediction error have been proposed by \cite{leatherman2017designing}. However, it has been recognized that Kennedy and O'Hagan model has severe identifiability issues \citep{tuo2015efficient,plumlee2017bayesian} and therefore, optimal designs based on their model can inherit similar problems. \cite{arendt2016preposterior} have proposed designing physical experiments to mitigate the identifiability issues in the Kenndy and O'Hagan model, but their procedure is computationally intensive. In this article, we will propose a much simpler approach to deal with the identifiability issue. \cite{ranjan2011follow} and \cite{williams2011batch} have proposed follow-up experimental designs for model calibration, but not an initial design that we plan to develop here.

The articles is organized as follows. In Section 2, we discuss the methodology of obtaining a robust experimental design for model calibration. In Section 3, we perform simulations on a toy example to illustrate the robustness of our proposed design to model discrepancy. In Section 4, we apply our design methodology to a problem relating to diaper line from the Procter \& Gamble (P\&G) company. We conclude the article with some remarks in Section 5.

\section{Robust Experimental Designs}

We will first explain how to develop experimental designs that are robust to model-form uncertainties. Then we will explain how to make them robust to parameter uncertainties. Finally we will explain how to incorporate computer model approximation uncertainties into this framework. These are now discussed in the following three subsections.

\subsection{Model-form uncertainties}
Let $y$ be the output of the physical system and $\bm{x} = \{x_1,..., x_p\}$ the set of inputs. Following \cite{kennedy2001bayesian}, we model the output as
\begin{equation}\label{eq:ko}
y = f(\bm{x}; \bm{\eta}) + \delta(\bm{x}) + \epsilon,
\end{equation}
where $f(\cdot;\cdot)$ is the computer model, $\bm{\eta} = \{\eta_1,..., \eta_q\}$ the set of unknown calibration parameters, $\delta(\bm x)$ the discrepancy function, and $\epsilon \stackrel{iid}\sim \mathcal{N}(0,\sigma^2)$ the random error. For the moment we will assume that the computer model is an easy-to-evaluate function or that the complex computer model has been replaced by an easy-to-evaluate surrogate model whose uncertainties can be neglected. Later we will see how such uncertainties can also be included in our approach.

As mentioned earlier, the model in (\ref{eq:ko}) has identifiability issues in the sense that for any value of $\bm \eta$ we can find a $\delta(\bm x)$ to get the same prediction \citep{tuo2015efficient,plumlee2017bayesian} and thus, $\bm \eta$ and $\delta(\bm x)$ cannot be estimated based on the data on $y$ alone unless some additional assumptions are imposed. Therefore, we will not directly use this model for developing the robust designs. We will put some belief in the computer model and hope that, if properly calibrated, we will not need the discrepancy term. Thus, we will first find an optimal design for estimating $\bm \eta$ ignoring the model discrepancy term. We will then separately find an optimal design for estimating $\delta(\bm x)$ and then integrate them together. This approach follows the sequential model building strategy proposed by \cite{joseph2009statistical}, where $\bm \eta$ is estimated by first ignoring the discrepancy. The discrepancy term is added only if it is necessary and if added, it is estimated by fixing $\bm \eta$ at its initial estimate. This mitigates the identifiability issues present in a joint estimation procedure.

Thus, first consider the case with no discrepancy, that is, $\delta(\bm x)=0$. Our aim is to choose a design to efficiently estimate $\bm \eta$ from the model
 \begin{equation}\label{eq:nobias}
y = f(\bm{x}; \bm{\eta}) + \epsilon.
\end{equation}
\cite{yang2010garza} showed that for a large class of non-linear models with moderate number of parameters, the de la Garza phenomena holds, i.e., for estimating $q$ unknown parameters, there exists an optimal design that contains exactly $q$ points. Thus, $\mathcal{D}_{\eta}=\{\bm x_1,\ldots,\bm x_q\}$. Suppose $\bm \eta_0$ is a guess value of $\bm \eta$. Then, a locally D-optimal design to estimate $\bm \eta$ can be obtained as
\begin{equation}\label{eq:local_util}
\mathcal{D}_{\eta} = \argmax_{\mathcal{D}\in \mathcal{X}} \left|\bm J_{0}^T\bm J_0\right|=\argmax_{\mathcal{D}\in \mathcal{X}} \left|\bm J_0\right|,
\end{equation}
where $\mathcal{X}$ is the experimental region and $\bm J_0$ is the sensitivity matrix evaluated at $\bm \eta=\bm \eta_0$, given by
\[\bm J_{0} =\begin{bmatrix}\frac{\partial f(\bm{x}_1;\bm{\eta})}{\partial \eta_1}& \cdots & \frac{\partial f(\bm{x}_1;\bm{\eta})}{\partial \eta_q} \\ \vdots && \vdots \\\frac{\partial f(\bm{x}_q;\bm{\eta})}{\partial \eta_1}& \cdots & \frac{\partial f(\bm{x}_q;\bm{\eta})}{\partial \eta_q}\end{bmatrix}_{\bm \eta=\bm \eta_0}.\]

Note that the optimal design depends on the guessed value of the unknown parameters when $f(\bm x;\bm\eta)$ is nonlinear with respect to $\bm \eta$. Therefore, the performance of the optimal design can deteriorate if the deviation of the guessed value from the truth and/or the nonlinearity increases. Of course the true value of $\bm \eta$ is unknown before the experiment and the uncertainties surrounding the guessed value should be incorporated into the design procedure. We will discuss about this issue in the next section.

Now that we have an optimal design to estimate the calibration parameters, we can focus our attention on the design to estimate the potential model discrepancy, $\delta(\bm x)$. Let us denote the design by $\mathcal{D}_{\delta}$. So the final design will be $\mathcal{D}= \mathcal{D}_{\eta} \bigcup \mathcal{D}_{\delta}$. \cite{kennedy2001bayesian} proposed to nonparametrically estimate $\delta(\bm x)$ using a Gaussian process. However, a Gaussian process model contains a set of unknown correlation parameters. These correlation parameters are nonlinear, which brings up the same issue as the calibration parameters. As before, we can make guessed values of the correlation parameters and try to develop locally optimal designs. However, there is something peculiar about these correlation parameters. The optimal design criteria such as the integrated mean squared error \citep{leatherman2017designing} are dominated by settings that produce low values of correlations \citep{joseph2019robust}. Thus, we only need to focus on a setting of the correlation parameters that minimizes the correlation. \cite{johnson1990minimax} have shown that in such limiting cases the D-optimal designs will reduce to maximin designs and G-optimal designs to minimax designs. In fact, these space-filling designs can be motivated using geometric considerations and are known to be robust to modeling choices. Since our objective is to develop model robust designs, they seem to be a perfect fit for our problem. Besides maximin and minimax, there are many choices for space-filling designs \citep{joseph2016space}. In this article, we will illustrate the methodology using maximum projection (MaxPro) designs \citep{joseph2015maximum}, but we will keep this choice flexible for the experimenter.

Now let us see how we can integrate the two designs $\mathcal{D}_{\eta}$ and $\mathcal{D}_{\delta}$. Since we plan to put some trust in the computer model, we will give some importance to the points in $\mathcal{D}_{\eta}$. Let $r$ be the number of replicates at these locations. Having a larger $r$ will improve the precision of the estimate of $\bm \eta$. In addition to that they will also help to get an unbiased estimate of $\sigma^2$. Suppose we have a budget for $n$ experiments. Then, we will use $qr$ runs for the estimation of $\bm \eta$. The remaining $n-qr$ runs will be used for estimating $\delta(\bm x)$. Those runs can be obtained using one of the space-filling designs mentioned before in such a way that they augment $\mathcal{D}_{\eta}$. 


The number of replications $r$ can be used to control the confidence we have in the computer model. If $r=0$, then we have no confidence in the computer model and our design will be purely based on the space-filling design. If $r=n/q$, then we have full confidence in the computer model, and all the runs will be allocated to efficiently estimating $\bm \eta$. In practice, before the experiments, we may not be able to accurately quantify the confidence in the computer model. In such a case, we recommend using $r=2$ as a default choice.

As a simple example, suppose the computer model is a linear model given by $f(\bm x;\bm \eta)=\eta_0+\eta_1x_1+\eta_2x_2+\eta_3x_1x_2$. D-optimal design for estimating this model is a $2^2$ full factorial design. Suppose we choose $r=2$ and we have a total budget for $n=13$ runs. The remaining $n-qr=5$ runs can be chosen to augment the eight runs using a space-filling criterion. For example, if  we use the MaxPro criterion, then the augmented design can be obtained sequentially by adding one point at a time using the \emph{MaxProAugment} function in the R package \emph{MaxPro} \citep{bamaxpro}.

\begin{figure}[h]
\begin{center}
\begin{tabular}{cc}
\includegraphics[scale=0.5]{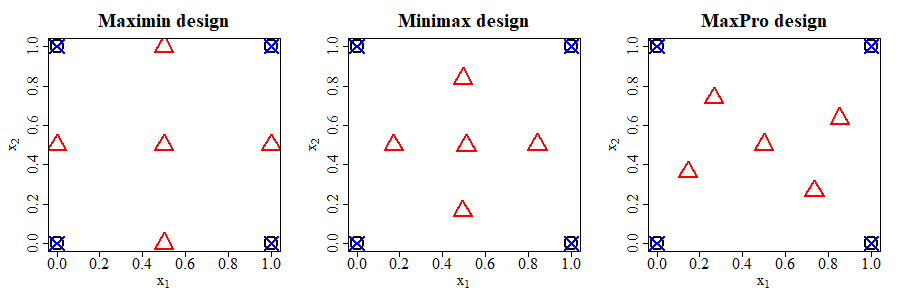}
\end{tabular}
\caption{Three model robust designs in 13 runs for estimating a linear model in two variables. The replicates of the optimal design points are shown as circles and crosses, and the space-filling  points as triangles.}
  \label{fig:linear model}
\end{center}
\end{figure}

Figure \ref{fig:linear model} (right) shows the 13-run design obtained using the Maxpro design for augmentation along with maximin (left) and minimax (center) augmentation strategies. We can see that all these designs will provide protection against possible departures from the linear model assumption. At first look, the maximin design seems most suitable for a physical experiment as it has fewer factor levels. However, in a high-dimensional space, this design may not give a good validation set as the points will occupy mostly the corners and boundaries of the hypercube. On the other hand, a MaxPro design simultaneously ensures space-fillingness in the full dimensional space as well as in all lower dimensional subspaces. Moreover, it can be easily extended to incorporate qualitative factors \citep{joseph2019designing}. However, a disadvantage of the MaxPro design is that it can produce too many levels for the factors, which can be inconvenient for a physical experiment. One quick remedy to this problem is to treat the factors as discrete-numeric with specified number of levels in the \emph{MaxProAugment} function \citep{bamaxpro}.


Note that we have used a balanced optimal design, where each point is replicated $r$ times. Further improvement in the optimal design can be achieved by using an unbalanced design. Let $w_i$ be the weight given to the $i$th point, where $\sum_{i=1}^{q}w_i=1$ and $w_i\ge 0$ for $i=1,\ldots,q$. Now both $\{\bm x_,\ldots,\bm x_{q}\}$ and $\{w_1,\ldots,w_{q}\}$ can be simultaneously obtained by maximizing the D-optimal criterion \citep{yang2013optimal}. Then the number of replications at each point can be obtained by rounding $rw_i$ to the nearest integer \citep{pukelsheim1992efficient} or using integer programming techniques \citep{sagnol2015computing}. For simplicity, we will use balanced optimal designs in this paper.

If we have some knowledge about the discrepancy function, we can further improve the design by including points optimal for estimating the discrepancy. For example, \cite{joseph2009statistical} found that a location-scale correction of the computer model can many times capture most of the discrepancy observed in the computer model. In other words, we are interested in fitting a model of the form
\begin{equation}\label{eq:JK}
y = f(\bm{x}; \bm{\eta}) + \beta_0+\beta_1f(\bm{x}; \bm{\eta})+\delta(\bm{x}) + \epsilon,
\end{equation}
where $\beta_0$ and $\beta_1$ are the adjustment parameters introduced to capture the location and scale bias in the computer model, respectively. Because $\beta_0$ and $\beta_1$ are linear parameters, they can be optimally estimated if we place two points where $f(\bm x;\bm \eta_0)$ is a maximum and a minimum over $\mathcal{X}$. That is,
\[\bm x_{max}=\argmax_{\bm x\in \mathcal{X}} f(\bm x;\bm \eta_0)\;\; \textrm{and}\;\; \bm x_{min}=\argmin_{\bm x\in \mathcal{X}} f(\bm x;\bm \eta_0).\]
These points can be directly added to the design. If they overlap with the D-optimal design points or are close to them, they act as additional replicates of those points. Adding these two points to the D-optimal design, we obtain a $(qr+2)$-point design. We will call this the location-scale corrected D-optimal design. We will augment this design, with an $(n-qr-2)$-point space-filling design, which is aimed at estimating $\delta(\bm x)$ in (\ref{eq:JK}). Thus, we obtain the desired $n$-point robust experimental design for model calibration.

\begin{figure}[h]
\begin{center}
\includegraphics[scale=0.35]{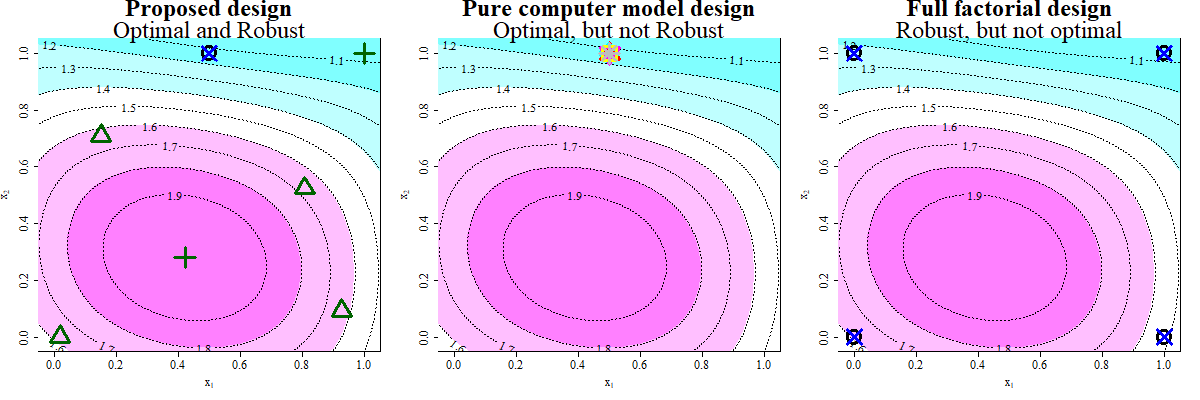}
\caption{Comparison of our proposed design with the `pure computer model' design, and full factorial design. The  replicates  of  the D-optimal design points are shown as circles and crosses, the maximum/minimum are shown as plus symbols, and the space-filling points as triangles.}
\label{fig:toy_comp}
\end{center}
\end{figure}

To illustrate the approach, we will use a toy example with two inputs and one calibration parameter. Suppose that we have a budget to perform eight physical experiments. The computer model is:
\[f(\bm x; \eta) = \exp\{-\eta(x_1-1.5x_2)^2\}+\exp\{-2\eta(x_1+x_2-0.7)^2\},\]
where $x_1,x_2 \in [0,1]^2$. Assume $\eta_0=0.5$. The one-point D-optimal design for estimating $\eta$ can be obtained as $\mathcal{D}_{\eta}=\{(0.50,1.00)\}$. Again, we replicate this point two times, that is $r=2$. Suppose we plan to model the discrepancy using (\ref{eq:JK}). Then the two points that maximize and minimize the computer model are obtained as $\bm x_3=(0.42,0.28)$ and $\bm x_4=(1.00,1.00)$, respectively. The remaining four runs are obtained by augmenting the location-scale corrected D-optimal design with the MaxPro design. The final model robust optimal design is shown in Figure \ref{fig:toy_comp} (left). For comparison, we have also shown a $2^2$ design with two replicates in Figure \ref{fig:toy_comp} (right), which would be a reasonable choice to make when we don't have a computer model. Furthermore, we also show the D-optimal design with eight replicates in Figure \ref{fig:toy_comp} (center). We call this design as the `pure computer model' design. 

If there is no model discrepancy, the `pure computer model' design is likely to provide the most accurate estimate of $\eta$. On the other hand, the full factorial design is likely to give the least accurate estimate of $\eta$. In the presence of model discrepancy, the `pure computer model' design is likely to perform poorly as it does not contain any space-filling points  for estimating model discrepancy. However, in both cases - presence or absence of model discrepancy, our proposed design, though may not be the best, is likely to provide a relatively accurate calibrated model as it contains both - the D-optimal design points optimal for estimating $\eta$, and the space-filling design points optimal for estimating the model discrepancy.

\subsection{Parameter uncertainties}
A weakness of the locally optimal designs is that the solution depends on the guessed value of $\bm \eta$. If the guessed value is not close to the (unknown) true value, the results may not be good especially when the model is highly nonlinear. As discussed in the introduction, a natural way to address these uncertainties is to use a Bayesian approach by placing a prior on $\bm \eta$. So let $p(\bm \eta)$ be the prior distribution. The $\bm \eta_0$ that we used in the previous section could be viewed as the mean or mode of this prior distribution. Instead of using a single value, now we will generate multiple values:
\[\bm \eta_i\sim p(\bm \eta),\;\; i=1,\ldots,m.\]
Following \cite{dror2006robust}, we will find local D-optimal designs for each of these values:
\begin{equation}\label{eq:local_utili}
\mathcal{D}_{\eta}^i = \argmax_{\mathcal{D}} \left|\bm J_i\right|,\;\; i=1,\ldots,m,
\end{equation}
where $\bm J_i$ is the sensitivity matrix evaluated at $\bm \eta=\bm \eta_i$. 

The set of local D-optimal designs consist of a total of $qm$ points. What we need is a $q$-point optimal design with each point replicated $r$ times. Here we will make a minor modification to better reflect the uncertainty in the optimal design points. We will find a $qr$-point optimal design and then combine the closest points to form replicates depending on the problem (cost of changing the set up, etc.). So here $r$ could be viewed more of as a confidence parameter instead of the number of replicates.

To do this, we start by replicating each of the $\mathcal{D}_{\eta}^i$'s $r$ times. This gives a total of $qmr$ points, which we need to reduce to $qr$ points. \cite{dror2006robust} proposed to use k-means clustering to reduce the set of local optimal designs to the optimal design robust to the parameter uncertainties. Instead, we will use the recently developed \emph{support points} \citep{mak2018support}, which better maintain the distribution of the local optimal designs.

We can also incorporate parameter uncertainty into finding the design points optimal for estimating the  location-scale correction parameters ($\beta_0,\beta_1$). To do this, we need to find the maximum and minimum of the computer model for each of the $m$ realizations of $\bm{\eta}$:
\[\bm x_{max}^{i}=\argmax_{\bm x\in \mathcal{X}} f(\bm x;\bm \eta_i)\;\; \textrm{and}\;\; \bm x_{min}^i=\argmin_{\bm x\in \mathcal{X}} f(\bm x;\bm \eta_i),\;\; i=1,\ldots,m.\]
We again use support points to obtain one point that represents the maxima and one for the minima. We add these two points to the $qr$ points to obtain the $(qr+2)$-point location-scale corrected D-optimal design. We will augment this design with  $(n-qr-2)$ runs of a space filling design, as done in Section 2.1. Thus, we obtain the desired $n$ run robust experimental design that incorporates parameter uncertainties.

As we need to find $\mathcal{D}_{\eta}^i$ and the maximum/minimum of $f(\bm x;\bm \eta_i)$ for $i=1,\ldots,m$, the procedure is much more computationally expensive than before. One idea to reduce the computational burden is to sample $\bm \eta_i$ using support points instead of random values from $p(\bm \eta)$, because support points will be able to represent the prior distribution with fewer points than a Monte Carlo sample. Thus we can use a much smaller $m$.

\begin{figure}[!htb]
\begin{center}
\includegraphics[scale=0.4]{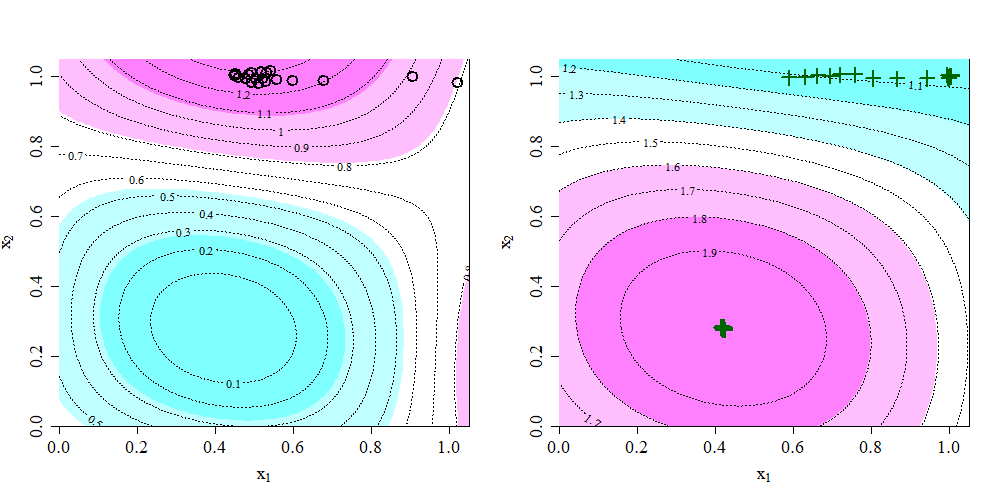}
\caption{(left): Local D-optimal design points, over the gradient contour of the computer model for $\eta = 0.5$; (right): Points corresponding to maximum and minimum of the computer model over the computer model contour for $\eta=0.5$.}
\label{fig:toy_paraUQ}
\end{center}
\end{figure}

Consider again the toy example with one calibration parameter used in the previous section. We assume that the calibration parameter has a prior distribution $\eta \sim \mathcal{N}(0.5,\,0.2^{2})\,.$ We obtain $m=20$ support points to represent this distribution using the R package \emph{support} \citep{supportR}. For each of the $m=20$ realizations, we obtain a one-point D-optimal design. The set of local D-optimal designs obtained are shown in Figure \ref{fig:toy_paraUQ} (left). Similarly, the $m=20$ points corresponding to maximum and minimum of the computer model are shown in Figure \ref{fig:toy_paraUQ} (right). 

\begin{figure}[h]
\begin{center}
\includegraphics[scale=0.4]{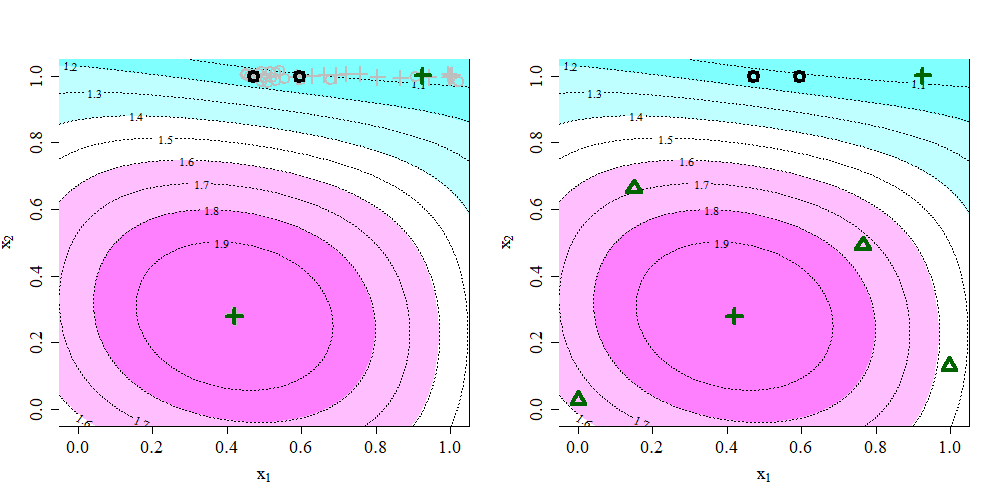}
\caption{(left): Four-run location-scale corrected D-optimal design; (right): Desired $n=8$ run robust experimental design; Both designs are shown over the computer model contour for $\eta = 0.5$; The circles correspond to the D-optimal design points, the plus symbols correspond to the maximum and minimum, and the triangles correspond to the space-filling design.}
\label{fig:toy_paraUQ2}
\end{center}
\end{figure}

We will reduce the $m=20$ local D-optimal design points to a two-run design using support points. Similarly, we will reduce the $m=20$ points corresponding to the maximum and minimum of the computer model to one point each. Figure \ref{fig:toy_paraUQ2} (left) shows the $qr+2 = 4$ runs thus obtained. We augment this four-run design with the Maxpro design to obtain the desired eight-run design as shown in Figure \ref{fig:toy_paraUQ2} (right). We observe that the design obtained is somewhat similar to the one obtained without incorporating parameter uncertainties in Figure \ref{fig:toy_comp} (left). However, unlike the design in Figure \ref{fig:toy_comp} (left), we observe that the two points of the D-optimal design for estimating $\eta$ are a bit farther apart (instead of overlapping). These two points can be combined to form two replicates at a single location, if needed.

\subsection{Surrogate model uncertainties}
So far we had assumed that $f(\bm x;\bm \eta)$ is an easy-to-evaluate model. In reality,  the computer model can be very expensive to evaluate. In such cases, first a computer experiment will be performed to obtain an approximation of $f(\bm x;\bm \eta)$. The approximate model is called a surrogate model or an emulator. Although there exist many different methods to obtain the surrogate model, Gaussian process modeling (or kriging) seems to be the most popular choice because of its ability to provide an uncertainty estimate \citep{sacks1989design}.

Let $\mathcal{S}$ denote the computer experimental design and $\bm y$ be the output values. Note that unlike in the physical experiment, the calibration parameters can be varied in the computer experiment. Thus, $\mathcal{S}$ has $p+q$ columns. For simplifying the notations, let $\bm u=(\bm x,\bm \eta)$ be the inputs in the computer experiment. Assume that $f(\cdot)$ is a realization of a Gaussian process:
\[f(\bm u)|\bm \eta\sim GP(\mu,C(\bm u;\cdot)),\]
where $\mu$ is the mean and $C(\bm u; \bm v)=Cov\{f(\bm u),f(\bm v)\}$ is the covariance function. See \cite{santner2018space} or \cite{gramacy2020surrogates} for details on Gaussian process modeling. Now, given the data, the posterior distribution of $f(\bm u)$ is also a Gaussian process given by
\begin{equation}\label{eq:posgp}
f(\bm u)|\bm \eta, \bm y\sim GP(\hat{f}(\bm u), C(\bm u;\cdot)-C(\bm u;\bm S)C^{-1}(\bm S;\bm S)C(\bm S;\cdot)),
\end{equation}
where $\hat{f}(\bm u)=\mu+C(\bm u;\bm S)C^{-1}(\bm S;\bm S)(\bm y-\mu\bm 1)$ is the surrogate model, $C(\bm u;\bm S)$ is the covariance vector with $i$th element $C(\bm u;\bm S_i)$, $C(\bm S;\bm S)$ is the covariance matrix, and $\bm 1$ is a vector of 1's.

Incorporating the surrogate model uncertainties into our design construction is conceptually very simple. We generate $m$ samples from $p(\bm \eta)$ and for each sample, we generate a realization of the function from (\ref{eq:posgp}):
\begin{eqnarray*}
\bm \eta_i&\sim & p(\bm \eta),\\
f_i(\bm u)|\bm \eta_i, \bm y &\sim & p(f(\bm u)|\bm \eta_i, \bm y)
\end{eqnarray*}
for $i=1,\ldots,m$. Now we can proceed to find the optimal design the same way as in the previous section except one difference. The gradients needed for the sensitivity matrix need to be calculated numerically for each new realization of $f(\cdot)$. This makes the procedure computationally very expensive.

It is a good idea to make the physical experiment design a subset of the computer experiment design, that is, the physical experiment design should be a nested design \citep{qian2009nested}. This avoids the confounding between the surrogate model approximation error and the discrepancy. To get a nested design, we have two options: (i) go back and run the computer simulations at the optimal physical experimental design points or (ii) choose the nearest points in $\mathcal{S}$ as the physical experimental design. If we use option (ii), there will be some loss of efficiency. Therefore, option (i) is preferred if the optimal design points are far away from $\mathcal{S}$ and it is feasible to run the computer simulation again. It is possible that the new simulations can change the surrogate model and hence the optimal design. So it seems some iterations will be needed to finalize the physical experimental design. However, we do not expect this to happen in most realistic cases unless there is too much uncertainty in the surrogate model approximation.

\begin{figure}[h]
\begin{center}
\includegraphics[scale=0.4]{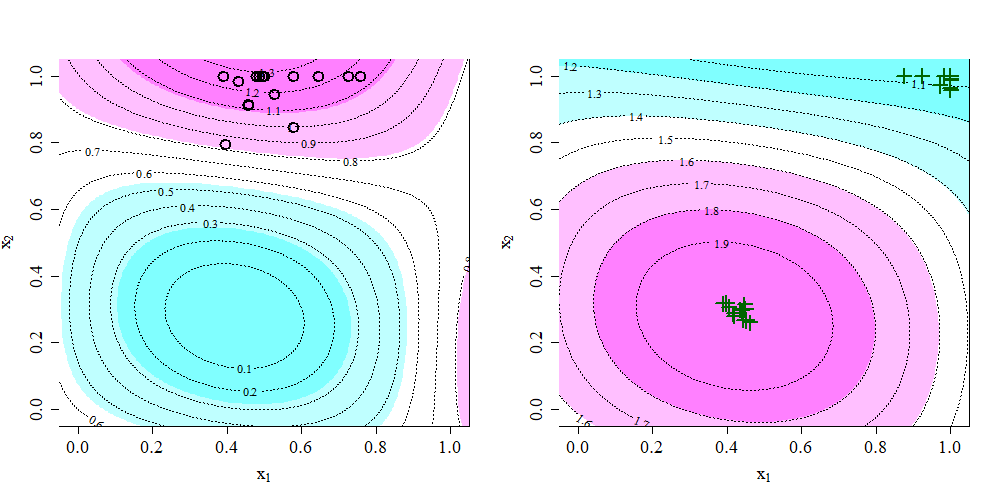}
\caption{(left): Local D-optimal design points, over the gradient contour of the computer model for $\eta = 0.5$; (right): Points corresponding to maximum and minimum of the computer model over the computer model contour for $\eta=0.5$.}
\label{fig:toy_surrUQ}
\end{center}
\end{figure}

Consider again the toy example. For illustrative purposes, we assume that the computer model is too expensive to compute. We consider a 30-run MaxPro design in $x_1, x_2,$ and $\eta$ to develop a surrogate Gaussian process model. Instead of the computer model $f(\bm x;\bm \eta)$, we use random realizations from the surrogate model $f_i(\bm u)|\bm \eta_i, \bm y$, for $i=1,\ldots,m$, to obtain the $m=20$ local D-optimal designs (Figure \ref{fig:toy_surrUQ} (left)), and also the $m=20$ maxima and minima (Figure \ref{fig:toy_surrUQ} (right)). We observe that due to uncertainty in the surrogate model prediction, the design points are more dispersed than in the case of no surrogate model uncertainty (Figure \ref{fig:toy_paraUQ}).

\begin{figure}[H]
\begin{center}
\includegraphics[scale=0.4]{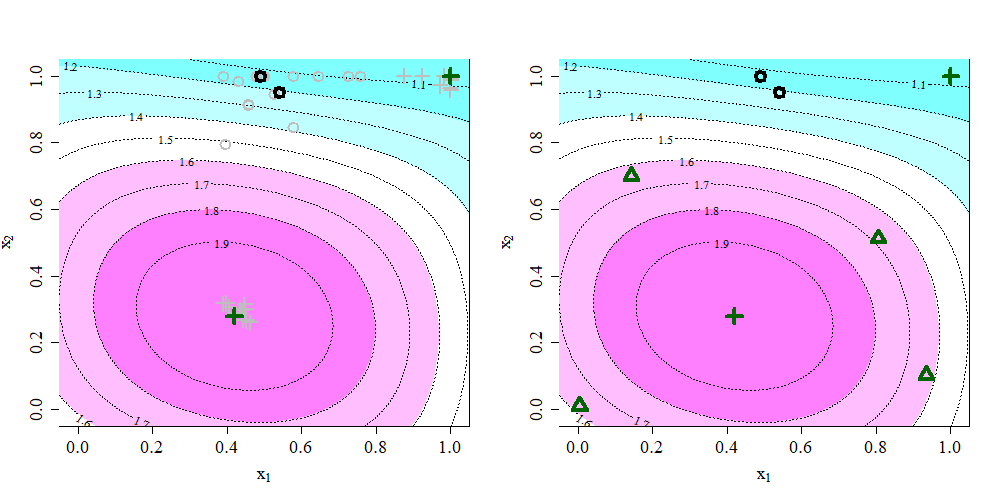}
\caption{(left): Four-run location-scale corrected D-optimal design; (right): Desired $n=8$-run robust experimental design; Both designs are shown over the computer model contour for $\eta = 0.5$; The circles correspond to the D-optimal design points, the plus symbols correspond to the maximum and minimum, and the triangles correspond to the space-filling design.}
\label{fig:toy_surrUQ2}
\end{center}
\end{figure}

We use support points to reduce the $qmr=40$ points of the local D-optimal designs to $qr=2$ points, and $m=20$ maxima and minima to one maximum and minimum, respectively. The resulting $qr+2$-run location-scale corrected D-optimal design obtained is shown in Figure \ref{fig:toy_surrUQ2} (left). We augment this design with an $n-qr-2=4$ run MaxPro design as shown in Figure \ref{fig:toy_surrUQ2} (right). We observe that the $n$ run design obtained is somewhat similar to the one without surrogate model uncertainty (Figure \ref{fig:toy_paraUQ2} (right)). However, the $qr=2$ points corresponding to the D-optimal design are scattered a little differently due to the surrogate model uncertainty.

\section{Simulations}

\begin{figure}[!htb]
\begin{center}
\includegraphics[scale=0.35]{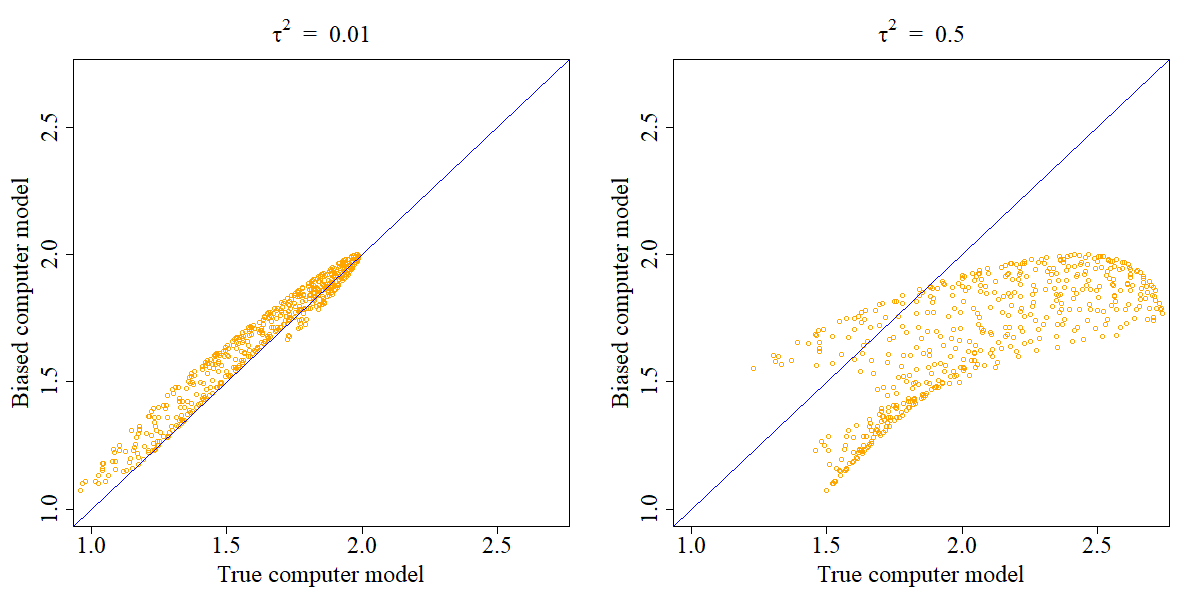}
\caption{Non-linear model discrepancy with increasing $\tau^2$.}
\label{fig:bias_visu}
\end{center}
\end{figure}

We illustrate the effectiveness of our proposed design in leveraging information from the computer model, while simultaneously being robust to potential model discrepancy. We consider the toy example again. However, we model the output as in \eqref{eq:ko}, which introduces a non-linear discrepancy in the computer model. Assume that $\delta(\bm{x})$ is a realization from a Gaussian process:
\begin{eqnarray*}
\delta(\bm{x}) \sim GP(0,\tau^2 R(\cdot)),
\end{eqnarray*}
where $R(\cdot)$ is the stationary correlation function and $\tau^2$ the variance. We wish to see the robustness of our proposed design with increasing magnitudes of non-linear model discrepancy. We increase the non-linear model discrepancy by increasing the value of $\tau^2$ (see Figure \ref{fig:bias_visu}).

We generate a random realization of $\delta(\bm{x})$, say $\delta_0(\bm{x})$ on all points of interest in the experimental design space $\mathcal{X}$ and obtain the physical experiment output at a point $\bm{x}$ as:
\begin{equation}
y(\bm{x}) = f(\bm{x};\eta) + \delta_0(\bm{x})+\epsilon,
\label{eq:toy_sim_output}
\end{equation}
where $\epsilon \stackrel{iid}{\sim} \mathcal{N}(0,0.05^2)$.
We simulate the physical experiment output on all points of the proposed design, and fit the model in \eqref{eq:JK} to the simulated output to estimate $\eta,\beta_0,\beta_1$ and $\delta(\bm{x})$. The estimation is done in two steps. First, we ignore $\delta(\bm{x})$, and find the posterior distribution of $\eta,\beta_0$ and $\beta_1$ using Markov chain Monte Carlo (MCMC) simulations. Using the posterior samples $(\eta^i,\beta_0^i,\beta_1^i)$, $i=1,\ldots,N$, we estimate the physical experiment output at a point $\bm{x}$ as:
\begin{equation}
\hat{f}(\bm{x}) = \frac{1}{N} \sum_{i=1}^N \left\{\beta_0^i+(1+\beta_1^i)*f(\bm{x};\eta^i)\right\}.
\label{eq:toy_estimate}
\end{equation}
The discrepancy at each of the design points can be obtained as $\delta_i=y(\bm x_i)-\hat{f}(\bm x_i)$ for $i=1,\ldots,n$. Let $\hat{\delta}(\bm{x})$ be the posterior mean of $\delta(\bm x)$ given $\mathcal{D}$ and $\bm{\delta}=(\delta_1,\ldots,\delta_n)'$. Then the prediction of the bias-corrected calibrated model at a point $\bm{x}$ is given by:
\begin{equation}
\hat{y}(\bm{x}) = \hat{f}(\bm{x})+ \hat{\delta}(\bm{x}). 
\label{eq:calib_pred}
\end{equation} 



We use \eqref{eq:calib_pred} to estimate the calibrated model output on a Sobol test dataset \citep{christophe2015randtoolbox} of 500 points. We use \eqref{eq:toy_sim_output} to simulate the physical experiment output on the same test dataset and compute the root mean square prediction error (RMSPE). We repeat the same exercise for the full factorial and the pure computer model designs. In order to evaluate the prediction performance of the calibrated model based on our proposed design, define the root mean square prediction error ratio (RMSPE) as:
\begin{equation}
\textrm{RMSPE ratio= } \frac{\vert \textrm{RMSPE}_{\it{competing \ design}}\vert}{\vert \textrm{RMSPE}_{\it{Proposed \ design}}\vert}.
\label{eq:RMSPE_ratio}
\end{equation}
If the RMSPE ratio is greater than 1, then we conclude that our proposed design performs better than the competing design. 

\begin{figure}[!htb]
\begin{center}
\includegraphics[scale=0.6]{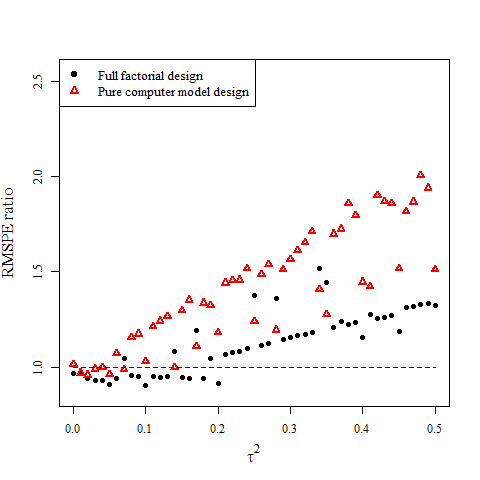}
\caption{Comparing performance of our proposed design with increasing non-linear model discrepancy.}
\label{fig:robustness}
\end{center}
\end{figure}

Figure \ref{fig:robustness} plots the RMSPE ratio with $\tau^2 = 0.0, 0.01,\ldots,0.5$. We observe that our proposed design performs comparably with other designs for low model discrepancy. However, for high model discrepancy, our proposed design performs better than both the other designs, and this performance gap increases with increasing discrepancy. This is because both the full factorial and the `pure computer model' designs lack the space-filling  points that are critical in estimating the non linear discrepancy. We observe that the performance of the `pure computer model' design deteriorates very rapidly with increasing discrepancy, which reinforces the importance of incorporating robustness in the design.

\section{A Real Example}

We apply our design strategy to a real example from P\&G. The model, based on first principles, involves a transformation of the P\&G diaper line, where absorbent gelling material is being applied to a substrate. The example has been slightly modified for the benefit of simplicity and to prevent disclosure of any potential sensitive information. The anonymized physics-based model is:

\begin{equation}
f(\bm{x};\bm{\eta}) = AGM*B*1000*2,
\label{eq:png_comp_model}
\end{equation}
where:
\[AGM =\bigg\{ \frac{(10^3x_2)^{2\eta_3}{x_5}^{2\eta_3-1}}{\eta_1^2(c_4)^{2\eta_3-1}} + 10^6k_2x_1x_2\bigg\}{\bigg\{(e^{x_4}-c_1)c_2+c_3\bigg\}x_5},\]
{\[B = 1-\bigg\{k_1-\Big(\frac{10^{-3}A}{(e^{x_4}-c_1)c_2+c_3}\Big)\frac{1}{x_5}-x_3\bigg\}^2\frac{1}{\eta_2},\]} 
where the controllable  process variables are $\bm{x}=\{x_1,...,x_5\}$ and the unknown calibration parameters $\bm{\eta}=\{\eta_1,...,\eta_3\}$. The details of the variables are omitted here due to confidentiality reasons. We assume that we have a budget of $n=16$ points to calibrate the physics-based model.

\begin{figure}[h]
\begin{center}
\includegraphics[scale=1]{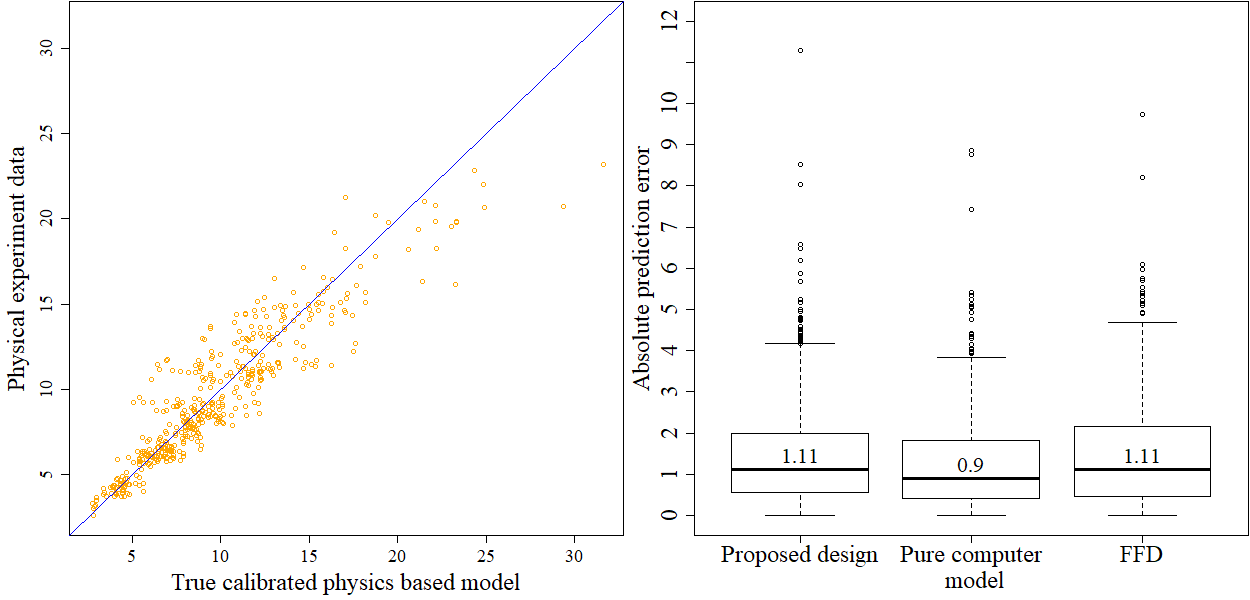}
\caption{(left): Physical experiment data vs the true calibrated physics-based model from P\&G; (right): Distribution of the absolute prediction error ratio in case of an unbiased physics-based model.}
\label{fig:png_nobias}
\end{center}
\end{figure}

P\&G has provided us with physical experiment data collected on 646 points. Of these, 463 data points are unique, and around 180 of them have one replicate each. The data were collected for multiple purposes including calibration of the physics-based model. The unknown values of the calibration parameters were estimated from this dataset, resulting in the calibrated model. Figure \ref{fig:png_nobias} (left) plots the physical experiment data against the model calibrated by P\&G. As the calibrated model seems unbiased, we will call this model the true calibrated model, and the estimated values of the calibration parameters the true calibration parameter values. We will use the true calibrated model to simulate physical experiment output for our proposed design, and other competing designs. If the true value of the calibration parameters is $\bm{\eta}_0$, and the true calibrated model is $f(\bm{x};\bm{\eta}_0$), then the physical experiment output can be simulated as:
\begin{equation}
y = f(\bm{x};\bm{\eta}_0)+\epsilon,
\label{eq:png_simulate_output}
\end{equation}
where $\epsilon \stackrel{iid}\sim \mathcal{N}(0,3)$. Note that the variance of noise is estimated from the physical experiment dataset.

In order to test our design methodology, we assume that the physics-based model mentioned in \eqref{eq:png_comp_model} has not been calibrated. This means that $\bm{\eta}$ is unknown, and that there may be model discrepancy. We will use our design methodology in Section 2 to propose a design for calibrating this model. First, we examine the uncertainties to be incorporated in our design. Since the physics-based model may be biased, we will incorporate model-form uncertainty in our proposed design. The calibration parameters  are assumed to have prior distributions: $\eta_1\sim\mathcal{U}[0.1,10], \eta_2\sim\mathcal{U}[0.1,1000],$ and $\eta_3\sim\mathcal{U}[0,1]$. Thus, we need to incorporate parameter uncertainty in our design. Since we have an easy-to-evaluate physics-based model, there is no need to consider surrogate model uncertainty in this example.

To incorporate information from the physics-based model, we include $r=2$ replicates of the D-optimal design in our proposed design. Since we have $q= 3$ calibration parameters, our design includes $qr= 6$ runs of the D-optimal design. To incorporate model-form uncertainty, we include two runs corresponding to maximum and minimum of the physics-based model for estimating the location-scale correction parameters, and $n-qr-2=8$ MaxPro design points for estimating the non-linear discrepancy. To incorporate parameter uncertainty, we use support points to generate $m=20$ realizations of the set of calibration parameters from their multivariate uniform prior distribution. This gives $qmr=120$ local D-optimal design points, which are reduced to $qr=6$ points using support points. Similarly, $m=20$ points corresponding to maxima and minima of the physics-based model are reduced to one point each. The remaining $16-qr-2=8$ runs are augmented by the MaxPro design.

We simulate the physical experiment output for our proposed design using \eqref{eq:png_simulate_output} and fit the model in \eqref{eq:JK} on the simulated output to estimate $\bm{\eta},\beta_0,\beta_1$, and $\delta(\bm{x})$. We follow the same method of estimation used  in Section 3. The calibrated model is given by \eqref{eq:calib_pred}. 

In order to compare the performance of our proposed design with other designs, we will similarly develop the calibrated model using those designs. We will use the 646-point P\&G test dataset to compare prediction performance of the calibrated model corresponding to our design with that of other designs. We compare our proposed design against (a) `pure computer model' design and (b) $2^{5-2}$ fractional factorial design (FFD) with two replicates. For generating the `pure computer model' design, we reduce the $qmr=120$ local D-optimal design points to $n=16$ runs using support points.  

We compute absolute prediction error on each of the 646 points in the test dataset using the calibrated model for each of the designs. Figure \ref{fig:png_nobias} (right) compares the distribution of absolute prediction error of competing designs - `pure computer model' design and FFD, with respect to our proposed design, over all the 646 points in the test dataset. We observe that the `pure computer model' design corresponds to the least prediction error. This is because there is no discrepancy in the physics-based model as seen in Figure \ref{fig:png_nobias} (left). Since all the 16 runs of the `pure computer model' design are based on D-optimal design points, it provides the most accurate estimate of the calibration parameters, and thereby the most accurate calibration. As our proposed design has six runs of the D-optimal design points, it is expected to do better than FFD. However, since FFD  happened to be close to the D-optimal design for this physics-based model, it gives a performance comparable to our proposed design.

\begin{figure}[!htb]
\begin{center}
\includegraphics[scale=1]{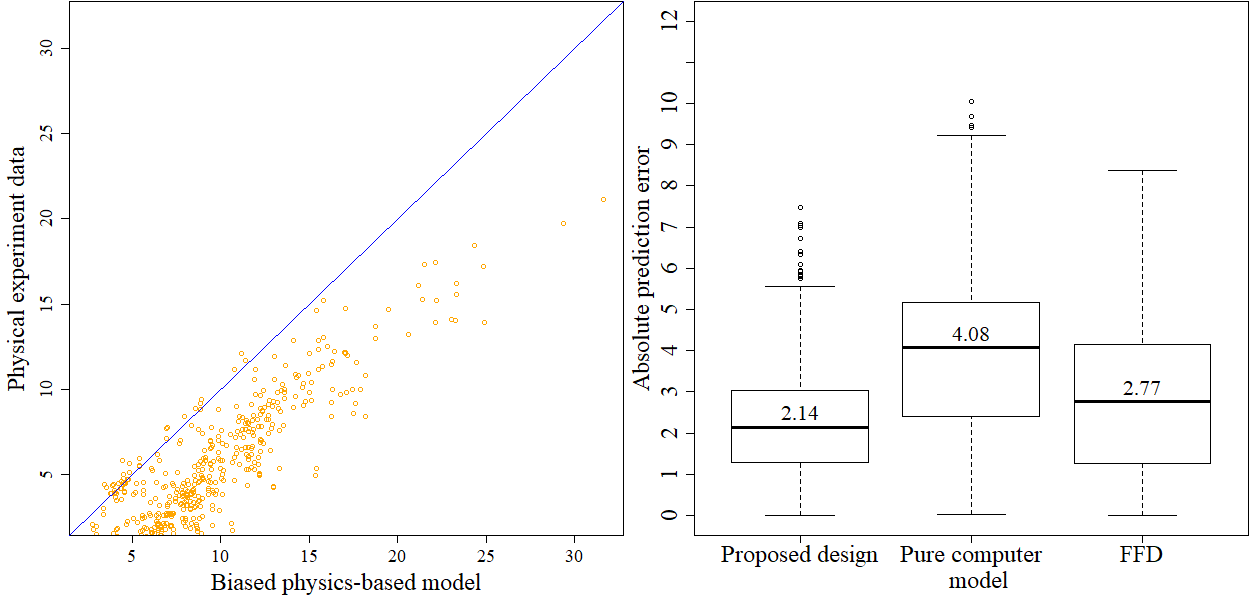}
\caption{(left): Physical experiment output vs the biased physics-based model output; (right): Distribution of the absolute prediction error ratio in case of a biased physics-based model.}
\label{png_bias}
\end{center}
\end{figure}

We are particularly interested to see the performance of our proposed design in the presence of model discrepancy. As the physics-based model $f(\bm{x};\bm{\eta})$ is unbiased, we will add a randomly generated realization of a Gaussian Process discrepancy to the true model to simulate the physical experiment output. Instead of simulating the physical experiment output using \eqref{eq:png_simulate_output}, we will simulate it as:
\begin{equation}
y = f(\bm{x};\bm{\eta}_0)+\delta_0(\bm{x})+\epsilon,
\label{eq:png_simulate_biased_output}
\end{equation}
where $\epsilon \stackrel{iid}\sim \mathcal{N}(0,3)$, and $\delta_0(\bm{x})$ is a random realization from $GP(0,\tau^2 R(\cdot))$
with $\tau^2=6$. Note that $\tau^2$ is chosen such that the discrepancy is large enough to be distinctly visible, see Figure \ref{png_bias} (left). 

Figure \ref{png_bias} (right) plots the absolute prediction error of the other designs with respect to our proposed design. We observe that our proposed design now corresponds to the least prediction error, which shows that it provides the most accurately calibrated model. A very drastic drop in the relative performance is observed in the case of `pure computer model' design, as it doesn't have optimal points to estimate model discrepancy. On the other hand, FFD does not use any information from the computer model. Thus, it does not have optimal points to estimate the calibration  parameters, which leads to a sub-optimal performance. In contrast, our proposed design uses the information from the computer model, while also accounting for model discrepancy, which leads to a better overall performance than both the competing designs. 

\section{Conclusion}

This article presented a strategy for designing physical experiments when a computer model is available to the experimenter. Optimal designs can be generated by directly using the computer model, but such designs are susceptible to possible model violations. Our design strategy augments the optimal design points with space-filling points. These space-filling points act as check points for the computer model and protect against possible model violations. The proposed designs can become inferior to the optimal designs when the computer model is perfect, but will be superior when the computer model is imperfect. Since the optimal designs is a subset of our proposed design, the loss of efficiency when the computer model is perfect is minimal and thus, we can claim our designs to be model robust.

We tried to make our design strategy simple and as flexible as possible. We also presented strategies to augment points when some information of the model discrepancy is available. For example, we showed that adding points corresponding to the maximum and minimum of the computer model can be used to efficiently estimate a location-scale bias of the computer model. These can be viewed as  ``features'' of the computer model that can act as useful model validation points. In fact, this also suggests that experimenters can extract other features from the computer model, such as all the local maxima and minima, and add them to the design, even when the connection to a discrepancy model is not evident. We hope to investigate this further in a future work.



\begin{center}
    {\Large\bf Acknowledgements}
\end{center}
This research is supported by the U.S. National Science Foundation grants DMS-1712642 and CMMI-1921646.
\bibliographystyle{ECA_jasa}
\bibliography{Ref_all}

\end{document}